\documentclass[aps,prl,reprint,twocolumn,superscriptaddress,showpacs,amsmath,amssym]{revtex4}

\usepackage{graphicx}
\usepackage{bm}

\usepackage{epsfig}

\RequirePackage{ifpdf}

\ifpdf
 \DeclareGraphicsExtensions{.jpg,.pdf,.jpeg}
\else
 \DeclareGraphicsExtensions{.eps,.ps,.eps.gz,.ps.gz}
\fi

\newcommand{\ppbar}{\ensuremath{p\bar{p}}}
\newcommand{\ttbar}{\ensuremath{t\bar{t}}}
\newcommand{\etal}{{\it et al.}}
\def \metnice {\ensuremath{{\not}{E_T}}}

\DeclareMathOperator{\JES}{JES}
\DeclareMathOperator{\DJES}{\Delta_{\JES}}

\DeclareMathOperator{\bg}{bg}

\begin{document}

\title{Top Quark Mass Measurement in the Lepton + Jets Channel Using a Matrix Element Method and {\it{in situ}} Jet Energy Calibration}

\affiliation{Institute of Physics, Academia Sinica, Taipei, Taiwan 11529, Republic of China} 
\affiliation{Argonne National Laboratory, Argonne, Illinois 60439, USA} 
\affiliation{University of Athens, 157 71 Athens, Greece} 
\affiliation{Institut de Fisica d'Altes Energies, Universitat Autonoma de Barcelona, E-08193, Bellaterra (Barcelona), Spain} 
\affiliation{Baylor University, Waco, Texas 76798, USA} 
\affiliation{Istituto Nazionale di Fisica Nucleare Bologna, $^z$University of Bologna, I-40127 Bologna, Italy} 
\affiliation{University of California, Davis, Davis, California 95616, USA} 
\affiliation{University of California, Los Angeles, Los Angeles, California 90024, USA} 
\affiliation{Instituto de Fisica de Cantabria, CSIC-University of Cantabria, 39005 Santander, Spain} 
\affiliation{Carnegie Mellon University, Pittsburgh, Pennsylvania 15213, USA} 
\affiliation{Enrico Fermi Institute, University of Chicago, Chicago, Illinois 60637, USA}
\affiliation{Comenius University, 842 48 Bratislava, Slovakia; Institute of Experimental Physics, 040 01 Kosice, Slovakia} 
\affiliation{Joint Institute for Nuclear Research, RU-141980 Dubna, Russia} 
\affiliation{Duke University, Durham, North Carolina 27708, USA} 
\affiliation{Fermi National Accelerator Laboratory, Batavia, Illinois 60510, USA} 
\affiliation{University of Florida, Gainesville, Florida 32611, USA} 
\affiliation{Laboratori Nazionali di Frascati, Istituto Nazionale di Fisica Nucleare, I-00044 Frascati, Italy} 
\affiliation{University of Geneva, CH-1211 Geneva 4, Switzerland} 
\affiliation{Glasgow University, Glasgow G12 8QQ, United Kingdom} 
\affiliation{Harvard University, Cambridge, Massachusetts 02138, USA} 
\affiliation{Division of High Energy Physics, Department of Physics, University of Helsinki and Helsinki Institute of Physics, FIN-00014, Helsinki, Finland} 
\affiliation{University of Illinois, Urbana, Illinois 61801, USA} 
\affiliation{The Johns Hopkins University, Baltimore, Maryland 21218, USA} 
\affiliation{Institut f\"{u}r Experimentelle Kernphysik, Karlsruhe Institute of Technology, D-76131 Karlsruhe, Germany} 
\affiliation{Center for High Energy Physics: Kyungpook National University, Daegu 702-701, Korea; Seoul National University, Seoul 151-742, Korea; Sungkyunkwan University, Suwon 440-746, Korea; Korea Institute of Science and Technology Information, Daejeon 305-806, Korea; Chonnam National University, Gwangju 500-757, Korea; Chonbuk National University, Jeonju 561-756, Korea} 
\affiliation{Ernest Orlando Lawrence Berkeley National Laboratory, Berkeley, California 94720, USA} 
\affiliation{University of Liverpool, Liverpool L69 7ZE, United Kingdom} 
\affiliation{University College London, London WC1E 6BT, United Kingdom} 
\affiliation{Centro de Investigaciones Energeticas Medioambientales y Tecnologicas, E-28040 Madrid, Spain} 
\affiliation{Massachusetts Institute of Technology, Cambridge, Massachusetts 02139, USA} 
\affiliation{Institute of Particle Physics: McGill University, Montr\'{e}al, Qu\'{e}bec, Canada H3A~2T8; Simon Fraser University, Burnaby, British Columbia, Canada V5A~1S6; University of Toronto, Toronto, Ontario, Canada M5S~1A7; and TRIUMF, Vancouver, British Columbia, Canada V6T~2A3} 
\affiliation{University of Michigan, Ann Arbor, Michigan 48109, USA} 
\affiliation{Michigan State University, East Lansing, Michigan 48824, USA}
\affiliation{Institution for Theoretical and Experimental Physics, ITEP, Moscow 117259, Russia}
\affiliation{University of New Mexico, Albuquerque, New Mexico 87131, USA} 
\affiliation{Northwestern University, Evanston, Illinois 60208, USA} 
\affiliation{The Ohio State University, Columbus, Ohio 43210, USA} 
\affiliation{Okayama University, Okayama 700-8530, Japan} 
\affiliation{Osaka City University, Osaka 588, Japan} 
\affiliation{University of Oxford, Oxford OX1 3RH, United Kingdom} 
\affiliation{Istituto Nazionale di Fisica Nucleare, Sezione di Padova-Trento, $^{aa}$University of Padova, I-35131 Padova, Italy} 
\affiliation{LPNHE, Universite Pierre et Marie Curie/IN2P3-CNRS, UMR7585, Paris, F-75252 France} 
\affiliation{University of Pennsylvania, Philadelphia, Pennsylvania 19104, USA}
\affiliation{Istituto Nazionale di Fisica Nucleare Pisa, $^{bb}$University of Pisa, $^{cc}$University of Siena and $^{dd}$Scuola Normale Superiore, I-56127 Pisa, Italy} 
\affiliation{University of Pittsburgh, Pittsburgh, Pennsylvania 15260, USA} 
\affiliation{Purdue University, West Lafayette, Indiana 47907, USA} 
\affiliation{University of Rochester, Rochester, New York 14627, USA} 
\affiliation{The Rockefeller University, New York, New York 10065, USA} 
\affiliation{Istituto Nazionale di Fisica Nucleare, Sezione di Roma 1, $^{ee}$Sapienza Universit\`{a} di Roma, I-00185 Roma, Italy} 

\affiliation{Rutgers University, Piscataway, New Jersey 08855, USA} 
\affiliation{Texas A\&M University, College Station, Texas 77843, USA} 
\affiliation{Texas Tech University, Lubbock, TX 79609, USA}
\affiliation{Istituto Nazionale di Fisica Nucleare Trieste/Udine, I-34100 Trieste, $^{ff}$University of Trieste/Udine, I-33100 Udine, Italy} 
\affiliation{University of Tsukuba, Tsukuba, Ibaraki 305, Japan} 
\affiliation{Tufts University, Medford, Massachusetts 02155, USA} 
\affiliation{University of Virginia, Charlottesville, VA  22906, USA}
\affiliation{Waseda University, Tokyo 169, Japan} 
\affiliation{Wayne State University, Detroit, Michigan 48201, USA} 
\affiliation{University of Wisconsin, Madison, Wisconsin 53706, USA} 
\affiliation{Yale University, New Haven, Connecticut 06520, USA} 
\author{T.~Aaltonen}
\affiliation{Division of High Energy Physics, Department of Physics, University of Helsinki and Helsinki Institute of Physics, FIN-00014, Helsinki, Finland}
\author{B.~\'{A}lvarez~Gonz\'{a}lez$^v$}
\affiliation{Instituto de Fisica de Cantabria, CSIC-University of Cantabria, 39005 Santander, Spain}
\author{S.~Amerio}
\affiliation{Istituto Nazionale di Fisica Nucleare, Sezione di Padova-Trento, $^{aa}$University of Padova, I-35131 Padova, Italy} 

\author{D.~Amidei}
\affiliation{University of Michigan, Ann Arbor, Michigan 48109, USA}
\author{A.~Anastassov}
\affiliation{Northwestern University, Evanston, Illinois 60208, USA}
\author{A.~Annovi}
\affiliation{Laboratori Nazionali di Frascati, Istituto Nazionale di Fisica Nucleare, I-00044 Frascati, Italy}
\author{J.~Antos}
\affiliation{Comenius University, 842 48 Bratislava, Slovakia; Institute of Experimental Physics, 040 01 Kosice, Slovakia}
\author{G.~Apollinari}
\affiliation{Fermi National Accelerator Laboratory, Batavia, Illinois 60510, USA}
\author{J.A.~Appel}
\affiliation{Fermi National Accelerator Laboratory, Batavia, Illinois 60510, USA}
\author{A.~Apresyan}
\affiliation{Purdue University, West Lafayette, Indiana 47907, USA}
% ...
\collaboration{CDF Collaboration\footnote{With visitors from $^a$University of Massachusetts Amherst, Amherst, Massachusetts 01003,
$^b$Istituto Nazionale di Fisica Nucleare, Sezione di Cagliari, 09042 Monserrato (Cagliari), Italy,
$^c$University of California Irvine, Irvine, CA  92697, 
$^d$University of California Santa Barbara, Santa Barbara, CA 93106
$^e$University of California Santa Cruz, Santa Cruz, CA  95064,
$^f$CERN,CH-1211 Geneva, Switzerland,
$^g$Cornell University, Ithaca, NY  14853, 
$^h$University of Cyprus, Nicosia CY-1678, Cyprus, 
$^i$University College Dublin, Dublin 4, Ireland,
$^j$University of Fukui, Fukui City, Fukui Prefecture, Japan 910-0017,
$^k$Universidad Iberoamericana, Mexico D.F., Mexico,
$^l$Iowa State University, Ames, IA  50011,
$^m$University of Iowa, Iowa City, IA  52242,
$^n$Kinki University, Higashi-Osaka City, Japan 577-8502,
$^o$Kansas State University, Manhattan, KS 66506,
$^p$University of Manchester, Manchester M13 9PL, England,
$^q$Queen Mary, University of London, London, E1 4NS, England,
$^r$Muons, Inc., Batavia, IL 60510,
$^s$Nagasaki Institute of Applied Science, Nagasaki, Japan, 
$^t$National Research Nuclear University, Moscow, Russia,
$^u$University of Notre Dame, Notre Dame, IN 46556,
$^v$Universidad de Oviedo, E-33007 Oviedo, Spain, 
$^w$Texas Tech University, Lubbock, TX  79609, 
$^x$Universidad Tecnica Federico Santa Maria, 110v Valparaiso, Chile,
$^y$Yarmouk University, Irbid 211-63, Jordan,
$^{gg}$On leave from J.~Stefan Institute, Ljubljana, Slovenia 
}}
\noaffiliation

\date{\today}

\begin{abstract}
A precision measurement of the top quark mass $m_t$ is obtained using
a sample of $\ttbar$ events from $\ppbar$ collisions at the Fermilab
Tevatron with the CDF II detector. Selected events require an electron
or muon, large missing transverse energy, and exactly four high-energy
jets, at least one of which is tagged as coming from a $b$ quark. A
likelihood is calculated using a matrix element method with
quasi--Monte Carlo integration taking into account finite detector
resolution and jet mass effects. The event likelihood is a function of
$m_t$ and a parameter $\DJES$ used to calibrate the jet energy scale
{\it{in situ}}. Using a total of 1087 events in 5.6 fb$^{-1}$ of
integrated luminosity, a value of $m_t = 173.0 \pm 1.2$ GeV/$c^2$ is
measured.
\end{abstract}


\maketitle


The top quark is the heaviest known fundamental particle in the
standard model of particle physics. Since the 1995 discovery of the
top quark at the Fermilab Tevatron~\cite{topdisc}, both the CDF and D0
experiments have been improving the measurement of its mass $m_t$,
which is a fundamental parameter in the standard model~\cite{TevEW}.
Loop corrections in electroweak theory relate $m_t$ (along with the
$W$ boson mass $m_W$) to the mass of the predicted Higgs boson. Thus,
precision measurements of $m_t$ help to constrain the value of the
Higgs boson mass~\cite{LEPcollab}.

This Letter describes the single most-precise measurement to date of
the top quark mass. It is performed on data collected by the CDF II
detector~\cite{cdfii} during Run II of the Fermilab Tevatron $\ppbar$
collider operating at $\sqrt{s} = 1.96$~TeV with a total integrated
luminosity of 5.6 fb$^{-1}$. The measurement is performed on candidate
$\ttbar$ events containing a lepton and four jets~\cite{mtm2_prd}. For
each event selected in this analysis, we calculate the probability of
observing that event by integrating the matrix element for $\ttbar$
production and decay over phase-space variables.  We use a neural
network discriminant to distinguish between signal and background
events to correct for the contribution due to background, and employ a
cut on the peak likelihood for a given event for additional rejection
of background and poorly modeled signal events.  This analysis offers
a gain of nearly 20\% in statistical precision over our previous
measurement~\cite{mtm2_prd} given an equal number of events; there, in
order to make the likelihood integration computationally tractable, we
introduced kinematic assumptions to reduce the dimensionality of the
integral. In the present analysis, we use a quasi--Monte Carlo
integration technique~\cite{qmc}, which converges more rapidly than
the typical $O(N^{-1/2})$ convergence of standard Monte Carlo
integration. This allows us to integrate over a total of 19 dimensions
in a computationally practical time, resulting in a more accurate
modeling of the event. Furthermore, in addition to the increased data
sample available with more integrated luminosity delivered by the
Tevatron, we have expanded our muon selection ability, which increases
the size of our data sample by nearly 30\%. In total, this measurement
improves our statistical precision by a factor of two over our
previous analysis with 1.9~fb$^{-1}$ ~\cite{mtm2_prd}.

In this measurement, the largest uncertainty is due to the uncertainty
in the jet energy scale (JES) determination. To reduce this
uncertainty, we calculate the likelihood as a two-dimensional function
of $m_t$ and a second parameter, $\DJES$, which corrects the jet
energies by a factor of $1 + \DJES \cdot
\sigma_j$, where $\sigma_j$ is the fractional
systematic uncertainty on the energy for a given
jet~\cite{jf,jetcorr}. The known $W$ boson mass is used to constrain
the $W \rightarrow q\bar{q'}$ decay, which yields information on the
$\DJES$ parameter. We can thus optimally combine events to reduce the
total uncertainty on $m_t$ due to JES.


Within the standard model, the top quark decays almost exclusively
into a $W$ boson and a $b$ quark. We define a ``lepton + jets'' event
as an event where one of the $W$ bosons produced by the $\ttbar$ pair
decays into a charged lepton (in this analysis, an electron or a muon)
and a neutrino, and the other into a $q\bar{q'}$ pair. The two $b$
quarks and two quarks from the $W$ boson then produce jets in the
detector~\cite{jetclu}. We thus require candidate events to have an
electron with $E_T > 20$~GeV or a muon with $p_T > 20$~GeV/$c$ in the
central detector ($|\eta| < 1$~\cite{eta_met}), or a muon with $p_T >
20$~GeV/$c$ obtained with a trigger~\cite{cdfii} on missing transverse
energy, $\metnice$~\cite{eta_met}, instead of a central muon. As the
neutrino energy is not detected, we require $\metnice > 20$~GeV in the
event. We also require exactly four jets with $E_T > 20$~GeV within
the region $|\eta| < 2.0$, at least one of which must be tagged as a
$b$ jet using a secondary vertex tagging algorithm~\cite{secvtx}. To
model $\ttbar$ events, we use Monte Carlo--simulated events generated
with the {\sc pythia}~\cite{pythia} generator for 15 different $m_t$
values ranging from 162 to 184 GeV/$c^2$.

Background events contributing to the selected sample are: a) events
in which a $W$ boson is produced in conjunction with heavy-flavor
quarks ($b\bar{b}$, $c\bar{c}$, or $c$); b) events in which a $W$
boson is produced along with light quarks, at least one of which is
mistagged as heavy flavor; c) QCD events that do not contain a true
$W$ boson; d) diboson ($WW$, $WZ$, or $ZZ$) or $Z$ + jets events; e)
single top events. We model the contribution from $W$+jets events
using {\sc alpgen}~\cite{alpgen}, single top events using {\sc
madgraph}~\cite{madgraph}, and diboson events with {\sc pythia}. The
$Z$+jets contributions are not modeled separately, but are included in
the $W$+light flavor contribution. All Monte Carlo samples are
processed with the CDF II detector response simulation
package~\cite{cdfsim}. The non-$W$ QCD background is modeled using a
sideband of data events selected to have a small contribution from
heavy boson decay. The numbers of background events are estimated with
the method used for the $\ttbar$ cross section
measurement~\cite{mtwoforu}, and are shown in
Table~\ref{table:back-ev}.

\begin{table}[htbp]
\caption{Expected sample composition for an integrated luminosity of
5.6 fb$^{-1}$. The $\ttbar$ contribution is estimated using a
  cross-section of 7.4 pb~\cite{m2xs} and $m_t$ = 172.5 GeV/$c^2$.} 
\label{table:back-ev}
\begin{center}
\begin{tabular}{l@{\hspace{0.2cm}}c@{\hspace{0.4cm}}c}
\hline \hline
Event type                & 1 $b$ tag          & $\ge$ 2 $b$ tags  \\
\hline
$W$+heavy flavor          & 129.5 $\pm$ 42.1   & 15.7 $\pm$ 5.5 \\
non-$W$ QCD               & 50.1 $\pm$ 25.5    & 5.5 $\pm$ 3.8 \\
$W$+light flavor mistag   & 48.5 $\pm$ 17.1    & 1.0 $\pm$ 0.4 \\
diboson ($WW$, $WZ$, $ZZ$)& 10.5 $\pm$ 1.1     & 1.0 $\pm$ 0.1 \\
Single top                & 13.3 $\pm$ 0.9     & 4.0  $\pm$ 0.4  \\
$Z \rightarrow \ell \ell$ + jets & 9.9 $\pm$ 1.2 & 0.8 $\pm$ 0.1 \\
Total background          & 261.8 $\pm$ 60.6   & 28.0  $\pm$ 9.6 \\
$\ttbar$ signal           & 767.3 $\pm$ 97.2   & 276.5 $\pm$ 43.0 \\
\hline
Total expected            & 1029.1 $\pm$ 114.5 & 304.5 $\pm$ 44.1 \\
\hline
Events observed           & 1016               & 247  \\ 
\hline \hline
\end{tabular}
\end{center}
\end{table}


For each event, we construct a likelihood as a function of $m_t$ and
$\DJES$ using the following integral:

\begin{equation}
  \begin{split}
\raisetag{72pt}
L(\vec{y} \mid m_t, \DJES) & = \frac{1}{N(m_t)} \frac {1}{A(m_t,\DJES)} \times \\
& \quad \sum_{i\,=1}^{24} w_{i} L_{i}(\vec{y} \mid m_t, \DJES) \\
L_{i}(\vec{y} \mid m_t, \DJES) & = \int \frac {f(z_1) f(z_2)}{FF}
~\textrm{TF}(\vec{y} \mid \vec{x}, \DJES) \times \\
& \quad |M (m_t,\vec{x})|^{2} ~d\Phi(\vec{x}), \\
\end{split}
\label{eq:perm_integral}
\end{equation}

\noindent
where $\vec{y}$ are the quantities measured in the detector (the
momenta of the jets and charged lepton), $\vec{x}$ are the
parton-level quantities that define the kinematics of the event,
$N(m_t)$ is a global normalization factor, ${A(m_t, \DJES)}$ is the
event acceptance as a function of $m_t$ and $\DJES$, $f(z_1)$ and
$f(z_2)$ are the parton distribution functions (PDFs) for incoming
parton momentum fractions $z_1$ and $z_2$, $FF$ is the relativistic
flux factor, TF$(\vec{y} \mid \vec{x}, \DJES)$ are the transfer
functions that describe the measured jet-momentum distributions given
the quark kinematics, $d\Phi(\vec{x})$ the phase space for the eight
particles in the $\ttbar$ production and decay process, and
$M(m_t,\vec{x})$ is the matrix element for the process. The integral
is calculated for each of the 24 possible permutations of jet-parton
assignment and then summed with weights $w_i$ determined by the
probability that a $b$ or light parton will result in a $b$-tagged or
untagged jet.

We use the Kleiss-Stirling matrix element~\cite{kleiss}, which is a
leading-order matrix element including both $q\bar{q} \rightarrow
t\bar{t}$ and $gg \rightarrow t\bar{t}$ production processes, as well
as all spin correlations. For the PDFs, we use the CTEQ5L
functions~\cite{cteq} for the incoming $q\bar{q}$ and gluons. The
normalization factor $N(m_t)$ is obtained by integrating the
Kleiss-Stirling matrix element with the PDFs and the flux
factor over the phase space formed by the two initial and the six
final-state particles. The acceptance $A(m_t,
\DJES)$ is obtained from simulated events where the parton directions
and momenta are smeared to simulate final-state jets.
The transfer functions connect the measured jets to the partons. We
construct the transfer functions by taking simulated $\ttbar
\rightarrow$ lepton + jets events in a wide range of masses and
matching the simulated jets to their parent partons. The transfer
functions are separated into momentum and angular terms; both are
constructed with dependence on the true jet $p_T$ and mass from the
Monte Carlo simulation. The transfer functions are constructed
separately for $b$ and light quarks, as well as for each of four bins
of jet $\eta$.
There are 32 phase space integration variables in
Eq.~(\ref{eq:perm_integral}) (for the two initial partons and six
final partons). Four of these are eliminated by energy and momentum
conservation, and four more by taking the charged lepton, neutrino,
and initial parton masses as known. In addition, we assume that the
lepton momentum is perfectly measured, and we neglect the effects of
the individual transverse momenta of the initial partons so that we
model only the transverse momentum of the total $\ttbar$ system, for
which we use a prior derived from Monte Carlo simulation. This leaves
a total of 19 dimensions over which the integral must be evaluated,
which we perform using a quasi--Monte Carlo technique.


Handling of background events is unchanged from our previous 
publication~\cite{mtm2_prd}.
We identify events likely to be background using a {\sc jetnet 3.5}
artificial neural network~\cite{jetnet} with ten inputs. We construct
distributions of the neural network output weight $u$ for signal,
$S(u)$, and background, $B(u)$, events, normalized to
their overall expected fractions, and calculate the expected
background fraction for a given event as $f_{\bg}(u) = B(u)/(B(u) +
S(u))$.

We calculate the likelihood for all candidate events under the
assumption that they are signal, but the combined likelihood contains
contributions from both signal and background events. However, only
the signal events contain information about $m_t$, so using Monte
Carlo--simulated events we compute the average likelihood for
background events and subtract it from the total likelihood:

\begin{multline}
\raisetag{36pt}
\log L_{\textrm{adj}}(m_t, \DJES) \\
= \sum_{i \in \textrm{events}}
  [\log L(\vec{y_i} | m_t, \DJES) - f_{bg}(u_i) \log
  \overline{L_{\textrm{bg}}}(m_t, \DJES)],
\label{eq:bg_formula}
\end{multline}

\noindent
where $L_{\textrm{adj}}$ is the adjusted total likelihood for a given
set of events, $L(\vec{y_i} | m_t, \DJES)$ is the likelihood for an
individual event from Eq.~(\ref{eq:perm_integral}), $f_{\bg}(u_i)$ is
the background fraction for a given event with a neural network output
$u_i$, and $\overline{L_{\textrm{bg}}}(m_t, \DJES)$ is the average
likelihood for a background event.

Besides background events, the sample includes events which contain a
real $\ttbar$, but where one or more of the four jets and/or the
lepton observed in the detector do not come directly from the
$t\bar{t}$ decay, and are not well-modeled by the signal likelihood or
handled by the background subtraction above. These events, which we
refer to as ``bad signal,'' have a variety of sources (extra jets from
gluon radiation, $\ttbar$ events where both $W$ bosons decay into
leptons or hadrons, $W \rightarrow \tau\nu$ decay, etc.) and make up
36\% of the simulated $\ttbar$ events for $m_t$ = 172.5~GeV/c$^2$. We
suppress these events by requiring that the peak log-likelihood value
for an event be at least 10. This cut retains 96.3\% of the
signal, while rejecting 30.8\% of the bad signal and 37.3\% of the
background.

\begin{figure}[bhtp]
\begin{center}
\centerline{
\epsfig{file=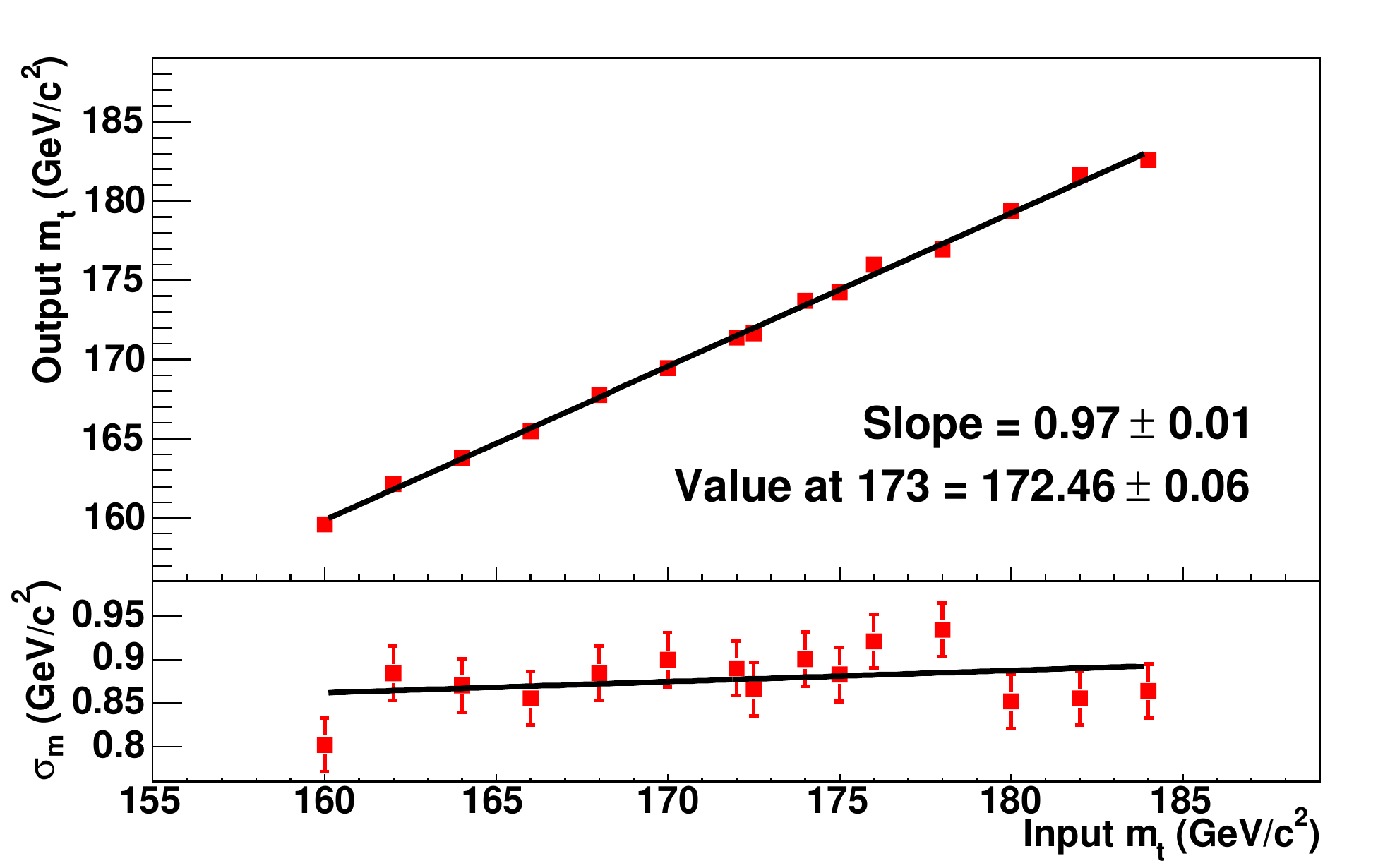,width=.45\textwidth}
}
\caption{Simulated experiment results using Monte Carlo signal and
  background events. Top: output $m_t$ vs. input $m_t$, before
  calibration is applied. Bottom: expected uncertainty $\sigma_m$
  vs. input $m_t$, with calibration applied. The lines are linear best
  fits.}
\label{fig:pe_results}
\end{center}
\end{figure}

We test and calibrate the method by constructing simulated experiments
using the Monte Carlo samples of $\ttbar$ events and background
described earlier. For a given input $m_t$ and $\DJES$, we perform
2000 experiments using a Poisson distribution with mean of 1089 events
(the number of events expected to pass the likelihood cut), and use
these to calibrate the measurement as a function of the input $m_t$
and $\DJES$. Figure~\ref{fig:pe_results} shows the output mass before
calibration and the calibrated expected uncertainty.

\begin{figure}[htbp]
\centerline{
\epsfig{file=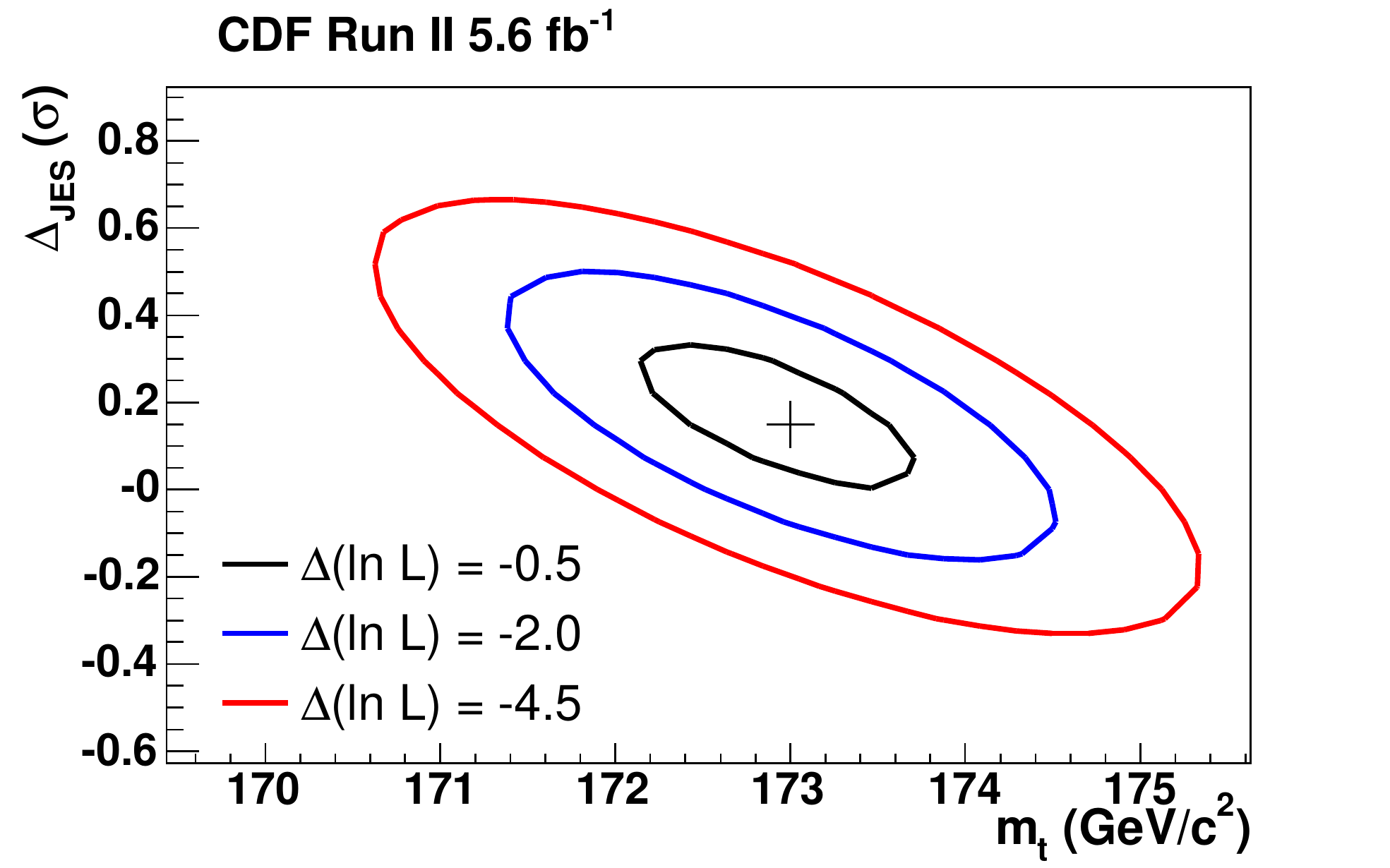,width=.45\textwidth}
}
\caption{Measured 2-D likelihood on the data events, with the
contours corresponding to a 1-$\sigma$, 2-$\sigma$, and 3-$\sigma$
 uncertainty in the final $m_t$ measurement from the profile
 method. The marker shows the point of maximum likelihood.
}
\label{fig:data_results}
\end{figure}

In the data we find a total of 1087 events which pass all of the
selection requirements (including the likelihood peak cut), of which
854 have 1 $b$ tag and 233 have $>$1 $b$
tag. Figure~\ref{fig:data_results} shows the resulting 2-D likelihood
contours for 1 $\sigma$, 2 $\sigma$, and 3 $\sigma$ after all
calibration.

To obtain a 1-D likelihood curve in $m_t$ only, we treat $\DJES$ as a
nuisance parameter and eliminate it using the profile likelihood
method~\cite{profile}, where we take the maximum value of the
likelihood along the $\DJES$ axis for each $m_t$ value.  The top quark
mass value extracted from the profile likelihood after calibration is
$m_t = 173.0 \pm 0.9$ GeV/$c^2$.  We can separate this uncertainty
into the statistical uncertainty on $m_t$ and the uncertainty due to
$\DJES$ by fixing the $\DJES$ value to its maximum likelihood
value. We find that the uncertainty from the resulting 1-D likelihood
is $0.7$ GeV/$c^2$, so we assign the remaining uncertainty of $0.6$
GeV/$c^2$ to $\DJES$ and conclude $m_t = 173.0 \pm
0.7~\textrm{(stat.)} \pm 0.6~\textrm{(JES)}~\textrm{GeV}/c^2$.

\begin{figure}[htbp]
\centerline{
\epsfig{file=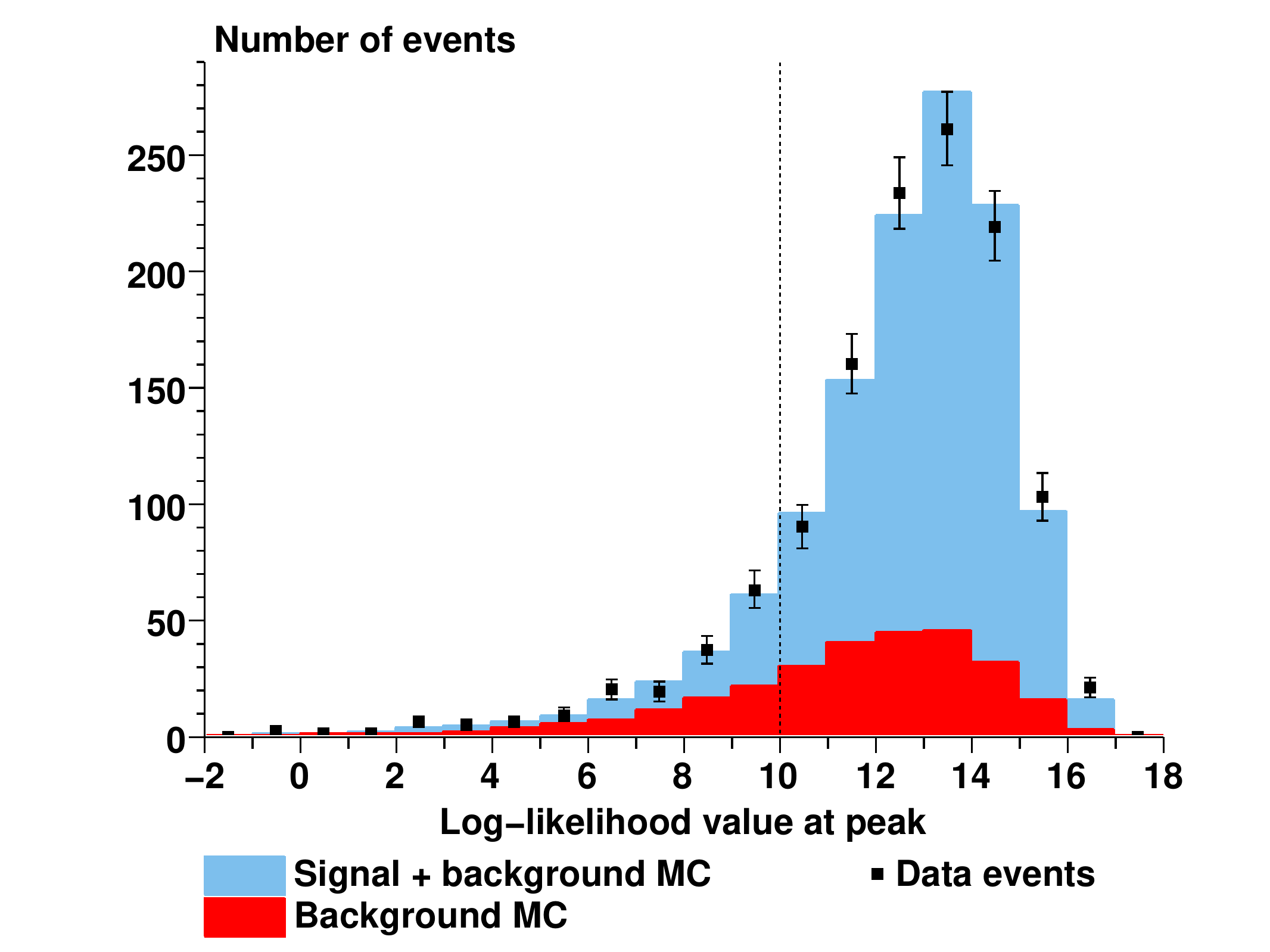,width=.45\textwidth}
}
\caption{Comparison of the log-likelihood value of the peak of
  likelihood curves for data and Monte Carlo events. The vertical line
  at 10 indicates the likelihood cut used in this analysis. A
  Kolmogorov-Smirnov test gives a confidence level of 0.93, showing
  good agreement between the two.}
\label{fig:likelihood_peaks}
\end{figure}

To validate the likelihood cut procedure,
we compare the peak values of the log-likelihood curves obtained with
data to those obtained with Monte Carlo--simulated events at $m_t =
172.5$ GeV/$c^2$ (the nearest available mass value). The results are
shown in Fig.~\ref{fig:likelihood_peaks}.

The systematic uncertainties on $m_t$, given in
Table~\ref{table:syst}, are derived using the methods described in
Ref.~\cite{mtm2_prd}. In brief, we include uncertainties coming from:
the calibration method; signal Monte Carlo modeling, evaluated by
comparing events simulated with the {\sc pythia} and {\sc
herwig}~\cite{herwig} generators; variations of the parameters used
for initial state radiation (ISR) and final state radiation (FSR); a
residual JES uncertainty because the JES uncertainty contains several
components with different $p_T$ and $\eta$ dependence; additional
uncertainties on the energy scale for $b$ jets; uncertainty on the
lepton $p_T$ scale; multiple hadron interactions, to take into account
uncertainty on the jet corrections as a function of the number of
interactions in the event; uncertainties arising from the PDFs used in
the integration; and the background modeling. This analysis includes a
systematic uncertainty due to color reconnection effects, not
considered in our previous analysis. We use {\sc pythia} version
6.4.20, which includes a color reconnection model~\cite{crref}, and
measure the difference between two tunes, Tune A, which is the tune
used in this analysis, and Tune ACR, which adds color reconnection
effects to Tune A. The individual systematic uncertainties are added
in quadrature to obtain the final total of 0.9 GeV/$c^2$.

\begin{table}[b!]
\caption{List of systematic uncertainties on $m_t$.}
\label{table:syst}
\begin{center}
\begin{tabular}{rc}
\hline \hline
Systematic source & Uncertainty (GeV/$c^2$) \\
\hline
Calibration                & 0.10 \\
MC generator               & 0.37 \\
ISR and FSR                & 0.15 \\
Residual JES               & 0.49 \\
$b$-JES                    & 0.26 \\
Lepton $p_T$               & 0.14 \\
Multiple hadron interactions & 0.10 \\
PDFs                       & 0.14 \\
Background modeling        & 0.33 \\
Color reconnection         & 0.37 \\
Total                      & 0.88 \\
\hline \hline
\end{tabular}
\end{center}
\end{table}


In conclusion, the measured top quark mass in a sample with 5.6
fb$^{-1}$ of integrated luminosity, with 1087 events passing all cuts,
is $m_t$ = 173.0 $\pm$ 0.7 (stat.) $\pm$ 0.6 (JES) $\pm$ 0.9 (syst.)
GeV/$c^2$, for a total uncertainty of 1.2 GeV/$c^2$. The improved
integration techniques and increased data sample make this the best
single measurement of the top quark mass to date, and it is comparable
in precision to the most recent combination for the top quark mass at
the Tevatron~\cite{TevEW}.

\begin{acknowledgments}
We thank the Fermilab staff and the technical staffs of the
participating institutions for their vital contributions. This work
was supported by the U.S. Department of Energy and National Science
Foundation; the Italian Istituto Nazionale di Fisica Nucleare; the
Ministry of Education, Culture, Sports, Science and Technology of
Japan; the Natural Sciences and Engineering Research Council of
Canada; the National Science Council of the Republic of China; the
Swiss National Science Foundation; the A.P. Sloan Foundation; the
Bundesministerium f\"ur Bildung und Forschung, Germany; the World
Class University Program, the National Research Foundation of Korea;
the Science and Technology Facilities Council and the Royal Society,
UK; the Institut National de Physique Nucleaire et Physique des
Particules/CNRS; the Russian Foundation for Basic Research; the
Ministerio de Ciencia e Innovaci\'{o}n, and Programa
Consolider-Ingenio 2010, Spain; the Slovak R\&D Agency; and the
Academy of Finland.
\end{acknowledgments}



\begin{thebibliography}{999}
\bibitem{eta_met} A particle's transverse momentum $p_T$ and tranverse
  energy $E_T$ are given by $|\vec{p}|\sin\theta$ and $E \sin \theta$
  respectively, where $\theta$ is the polar angle with respect to the
  proton direction (z-axis). The pseudorapidity $\eta$ of a particle's
  three-momentum is defined by $\eta = -\ln(\tan(\theta/2))$. The
  missing $E_T$, $\metnice$, is defined by $ \vec{\metnice} =
  -|\sum_{i}{E_{T_i} {\hat{n}_{T_i}}} |$, where $\hat{n}_{T_i}$ is the
  unit vector in the $x$--$y$ plane pointing from the primary vertex
  to a given calorimeter tower $\it{i}$, and $E_{T_i}$ is the $E_T$
  measured in that tower.
\end{thebibliography}
\end{document}